\documentclass[superscriptaddress,reprint,10pt,a4paper]{revtex4-2}

\usepackage{bm}

\usepackage{amsmath}
\usepackage{siunitx}
\DeclareSIUnit{\angstrom}{\textup{\r A}}
\let\ts\textsubscript
\usepackage{graphicx}
\usepackage{xcolor}
\usepackage{makecell}
\usepackage[normalem]{ulem}

\newcommand{\textsub}[2]{{#1}_{\text{#2}}}
\newcommand{\EF}{E_{\text F}}
\newcommand{\MgGaic}{(MgGa)\ts{ic}}

\newcommand{\ii}{\mathrm{i}}
\newcommand{\dd}{\mathrm{d}}
\newcommand{\ee}{\mathrm{e}}
\DeclareMathOperator{\erf}{erf}

\newcommand{\EQ}[1]{Eq.~(\ref{#1})}
\newcommand{\EQs}[1]{Eqs.~(\ref{#1})}
\newcommand{\FIG}[1]{Fig.~\ref{#1}}

\newcommand{\TBL}[1]{Table~\ref{#1}}

\begin{document}

\title{Multistability of interstitial magnesium and its carrier recombined migration in gallium nitride}

\author{Yuansheng Zhao}
\email{yszhao@g.ecc.u-tokyo.ac.jp}
\affiliation{Institute of Materials and Systems for Sustainability, Nagoya University, Nagoya 464-8603, Japan}
\affiliation{Quemix Inc., Chuo-ku, Tokyo 103-0027, Japan}
\affiliation{Department of Physics, The University of Tokyo, Bunkyo-ku, Tokyo 113-0033, Japan}

\author{Kenji Shiraishi}
\affiliation{Institute of Materials and Systems for Sustainability, Nagoya University, Nagoya 464-8603, Japan}
\affiliation{Center for Innovative Integrated Electronic Systems, Tohoku University, Sendai 980-8572, Japan}
\affiliation{Graduate School of Engineering, Nagoya University, Nagoya 464-8601, Japan}

\author{Tetsuo Narita}
\affiliation{Toyota Central R\&D Laboratories, Inc., Nagakute, Aichi 480-1192, Japan}

\author{Atsushi Oshiyama}
\affiliation{Institute of Materials and Systems for Sustainability, Nagoya University, Nagoya 464-8603, Japan}
\affiliation{Center for Innovative Integrated Electronic Systems, Tohoku University, Sendai 980-8572, Japan}

\begin{abstract}
We present density-functional-theory calculations which provide a microscopic 
picture of the recombination-enhanced migration of interstitial Mg in GaN. We 
determine stable structures and migration pathways with accurate HSE 
approximation to the exchange-correlation energy, and also computed 
recombination rates using the obtained energy spectrum and wavefunctions. It is  
found that the migration between the most stable octahedral sites (Mg\ts O) via newly found 
interstitial complex structure shows the lowest migration energy in which one or 
two electrons are captured during the migration, that the most stable charge state 
of $2+$ changes to $1+$ or neutral, and that by this recombination of carriers the 
migration barrier is significantly reduced. Starting from Mg\ts O$^{2+}$, Mg 
captures an electron becoming the $1+$ charge state and overcomes the barrier of 
\SI{1.65}{eV}, much reduced from \SI{2.23}{eV} in case of the migration with the $2+$ charge 
state kept. Moreover, further electron capture is realized accompanied by 
substantial structural relaxation, thus Mg becoming neutral. Detailed HSE 
calculations for this second capture show that the migration barrier is \SI{1.55}{eV}, thus 
clarifying the important role of the carrier recombination for Mg migration in GaN. 
These findings are corroborated by the present quantitative calculations of 
recombination rates based on electronic Hamiltonian constructed from our DFT-obtained energy spectrum. 
The timescale of the recombination is clarified to be in or under the timescale 
of the migration with typical electron density and the enhancement is expected to be significant.
\end{abstract}

\maketitle

\section{Introduction}

Gallium nitride (GaN), already a premier semiconductor in optoelectronics,
is now also emerging as a promising material for advancing future power electronics thanks to its high breakdown voltage, 
high carrier mobility and resistance under harsh conditions. 
In the fabrication of such semiconductor devices,
doping of electron donor and acceptor impurities to generate electrons and holes carrying electric current is always an essential topic
\cite{sze2021physics,narita2020characterization,pantelides1978electronic}.
As for GaN, an Mg atom substitutes for the Ga site and works as an acceptor exclusively.
In the process of manufacturing planar optical devices, 
the Mg has been usually doped during the epitaxial growth of GaN \cite{akasaki2015nobel,amano2015nobel,nakamura2015nobel}.
However, in the case of power devices with more complicated structures, to achieve the desired impurity concentration profiles, 
ion implantation followed by thermal annealing is indispensable.
Unfortunately, the Mg doping via such method has not been successfully achieved before due to difficulties encountered during annealing such as N desorption. 
Recently, Sakurai and collaborators at Nagoya University have succeeded in forging $p$-type GaN through Mg implantation followed by annealing under high pressure of nitrogen \cite{sakurai2019highly,sakurai2020redistribution},
advancing the doping technology in GaN to a next stage comparable to Si technology.

During the annealing process, 
the diffusion and migration of the defects, both intrinsic and extrinsic, generated by 
the ion implantation process are key phenomena which govern the profiles of donor and acceptor concentrations. 
As stated above, Mg located at the Ga site is an exclusive candidate for acceptors in GaN and clarification of its diffusing properties are crucial to fabricate GaN devices. 
In the diffusion processes, the incorporated Mg atoms in GaN migrates with the aid of intrinsic defects such as vacancies or interstitials. 
Under Mg implantation, the density of such intrinsic defects generally increases.
In particular, the Mg interstitial atoms are expected to be a major diffusing species.
Hence, at this point, in order to achieve better device properties, it is highly important to unveil the migration mechanism of 
the interstitial Mg, which is one of the most important processes under Mg implantation.

Very recently, we reported in a previous Letter \cite{zhao2024first} that the interstitial Mg exhibits complex local atomic structures
and rich migration behaviors:
The obtained total-energy landscape shows that the migration pathway with the lowest energy barrier involves the global energy minimum,
Mg at octahedral (O) site (Mg\ts O), and the lowest metastable configuration labeled as \MgGaic.
Most importantly, 
we proposed that Mg migrates through the so-called recombination-enhanced mechanism \cite{baraff1983theory,pantelides1984theory,pantelides1984microscopic},
in which the charge state of Mg is varying.
It is found that this phenomenon greatly reduces the migration barrier by a half \si{eV} compared with the migration with the fixed charge state of $2+$ \cite{miceli2017migration}.
This can well explain the small values of migration barrier reported in some experiments, especially in $n$-type GaN \cite{pan1999doping}, as well as the suppression of migration during high-pressure N\ts 2 annealing \cite{sumida2021effect}.

The carrier recombination-assisted migration is closely related to the electronic structure near the gap since carriers should be localized by being trapped at the particular levels in the case. 
Such localization of the carriers is naturally accompanied by structural relaxation which in turn affects the electronic structure.  
Additionally, the rate of such carrier capture clearly plays an important role in determining the effectiveness of such migration processes since the enhancement can only be significant if the capture rate is high enough compared with the time scale of migration itself. 
It has been generally argued that the rate of recombination is crucial for explaining the migration of Al in Si, where both the energy barrier and pre-exponential factor are affected by the Fermi level position \cite{pantelides1984theory}.
Hence the recombination enhanced/retarded migration is a complex phenomenon, which has not been addressed microscopically in the past and 
also in our previous report \cite{pantelides1984theory,zhao2024first}. 
In this paper, we aim to provide a comprehensive discussion about the carrier recombination process during the migration of interstitial Mg, including charge state dependence of the migration pathway and recombination rate.
In addition, we provide a validation of the density functional theory (DFT) \cite{kohn1965self,hohenberg1964inhomogeneous} calculation scheme of structural optimization with generalized gradient approximation (GGA) \cite{perdew1996generalized} used in the previous Letter by performing hybrid-functional calculations in this paper.

\section{Calculation details}

\subsection{DFT calculation}\label{sec:dft}

The first-principles calculation is performed based on standard DFT scheme implemented in the VASP \cite{kresse1996efficient}
with the similar conditions as our previous report \cite{zhao2024first}.
We use a projector augmented wave (PAW) potential \cite{blochl1994projector,kresse1999ultrasoft} and expand the electron orbitals using the plane-wave basis set with a cutoff energy of \SI{410}{eV}.
The HSE hybrid functional \cite{heyd2003hybrid} is employed for the exchange-correlation (XC) energy with the mixing ratio $\alpha=0.29$ of Fock exchange, which can reproduce the band gap $\SI{3.4}{eV}$ of wurtzite GaN as well as the lattice constant $a=b=\SI{3.19}{\angstrom}$ and $c=\SI{5.19}{\angstrom}$ \footnote{In our previous letter \cite{zhao2024first}, we have used the lattice constant optimized at GGA level, resulting a different value for the mixing ratio for Fock exchange to be employed to produce the correct band gap.}.
A defect is embedded in a $4\times4\times3$ supercell consisting of 192 lattice sites in total and relaxed until the force on each atom becomes less than \SI{0.02}{eV/\angstrom}. 
The Brillouin zone is sampled using a single $k$ point located at $(1/2,1/2,1/2)$.
To examine the validity of the $k$ point sampling, the effect of Ga $3d$ orbital, supercell volume and also the GGA-HSE difference, we have performed the formation-energy calculations with various calculational parameters by taking the most stable interstitial Mg complex, \MgGaic{} (see \FIG{fg:path} (b) and Fig.~1 of Ref.~\cite{zhao2024first}), as an example.
These careful examinations are excerpted in \TBL{tb:ergs}. 
It is found that while the absolute total energy is not converged with the single $k$ point (compared with calculations with $2\times2\times2$ $k$-point mesh), the energy difference is reproduced within \SI{0.1}{eV}. 

Additionally, this choice of $k$ point has another advantage: For the calculation of the recombination rate, the total energies near the crossing point of the two energy surfaces in which the defect-origin state is either occupied or unoccupied are required (see below).
In the present case, the orbital energy of such localized defect state is close to the conduction band minimum (CBM) which is located at $\Gamma$ point. Hence with $\Gamma$ point sampling, the CBM character is exaggerated wherever the defects state is located near the CBM during the total-energy analyses. On the other hand, the $(1/2,1/2,1/2)$ sampling is free from such exaggeration becausue there is no band state at CBM.

During the whole calculation, we have put Ga $3d$ orbitals in the core state and not explicitly treated since this only results in a negligible difference in formation energies.

Regarding the supercell volume, we have relaxed the supercell volume containing an additional Mg atom and found that the total energy is reduced by \SI{0.17}{eV}, which is acceptable for the present purpose.

\begin{table}
    \caption{Formation energies of neutral Mg interstitial complex \MgGaic{} under various optimization conditions. ``Single $k$ point'' denotes $k$-point sampling using only $(1/2,1/2,1/2)$.
All the formation energies are calculated by HSE and $2\times2\times2$ $k$ points unless explicitly stated.}
    \label{tb:ergs}
    \begin{tabular}{ll}
    \hline\hline
    \makecell[l]{Optimization condition\\(XC and $k$ points, etc.)}  & {Formation energy (eV)} \\ 
    \hline
    HSE, $2\times2\times2$ $k$ points & \hphantom{$-$}0 (as the zero point)\\
    \hline
    HSE, single $k$ point & \makecell[l]{\hphantom{$-$}0.00 ($2\times2\times2$ $k$ points)
\\\hphantom{$-$}0.05 (single $k$ point)}\\
    \hline
    \makecell[l]{HSE, single $k$ point\\Ga 3$d$ included} & $-0.03$ (single $k$ point)\\
    \hline
    \makecell[l]{HSE, single $k$ point\\Cell volume relaxation} & $-0.17$\\
    \hline
    GGA, $2\times2\times2$ $k$ points & \hphantom{$-$}0.08\\
    \hline
    GGA, single $k$ point & \hphantom{$-$}0.04\\
    \hline
    \hline
    \end{tabular}
    \end{table}

The formation energy of the structure $\zeta$, with charge state $q$ and the Fermi level $\EF$, is computed as 
\begin{equation}\label{eq:form-erg}\begin{split}
E(\zeta,q,\EF)&{}=\textsub E{tot}(\zeta,q)-\left(\textsub E{ref}+\sum_\alpha \mu_\alpha \Delta N_\alpha\right)\\
&{}+q(\textsub EV+\EF)+\textsub E{corr}.
\end{split}\end{equation}
Here, $\textsub E{tot}(\zeta,q)$ and $\textsub E{ref}$ represent the total energies of the structure $\zeta$ (i.e., the supercell model of the defects) and the reference structure (perfect GaN supercell), respectively.
$\Delta N_\alpha$ is the number difference in element $\alpha$ between 
the structure $\zeta$ and the reference, and the element $\alpha$ has chemical potential $\mu_\alpha$.
$\textsub EV$ is the energy of valence band maximum (VBM) which is used as the zero point of $\EF$.
Finally, $\textsub E{corr}$ is the correction term arising from the finite size of the supercell \cite{freysoldt2009fully}.

For determination of the migration pathways between stable or metastable configurations and their energy barriers, 
we have first used the climbing image nudged elastic band (CI-NEB) \cite{henkelman2000climbing} method to locate the transition state (i.e., the saddle point).
Other structures during the migration are obtained by descending from the transition point,
i.e., optimizing structures $\bm x_I(\xi)$ constrained on a set of the hyperplanes with fixed distance to the transition state
\[\sum_I^{\text{all atoms}} |\bm x_I(\xi)-\bm x_I^0|^2=\xi^2,\]
with $\bm x_I$ and $\bm x_I^0$ denoting the coordinates of atom $I$ in the structure to be optimized and in the transition state, respectively.
By varying the value for the \emph{migration coordinate} $\xi$, we can obtain the full migration pathway, parametrized by the variable $\xi$.

\subsection{Recombination rates}\label{sec:cap}

\subsubsection{Overview}

We here evaluate the rate of electron capture during the migration, i.e., the nonradiative carrier recombination due to the electron-lattice (phonon) interaction. 
Such carrier capture is generally promoted when the two electron states, e.g., the extended conduction band (CB) state and the localized defect state in the current case, come close in energy 
so that the energy-conservation law is readily satisfied.
Therefore, 
to facilitate the carrier recombination, the migrating atom 
first move to the crossing point of the 
total-energy surfaces of the two charge states in question, which will raise the total energy. 
In the present work, this energy increase is
fully considered by the total energy calculation along 
the migration pathways (see next section).

Far from the crossing point, the electronic wavefunction evolves nearly adiabatically and follows the motion of ions.
However, the adiabatic approximation breaks down near the crossing and an electron initially on CB can be captured to the defect level.
The probability of such capture is an
important ingredient for the recombination rates, and may be calculated using several methods for deep defects (without migration) \cite{henry1977nonradiative,shi2012ab,barmparis2015theory,goguenheim1990theoretical,gutsche1982non,alkauskas2014first}.
For such calculation in the current case (during the migration), we describe the migration itself classically similar to Ref.~\cite{henry1977nonradiative}, in accord with describing the migration path classically using $\bm x_I(\xi)$.
Thus, the migration coordinate $\xi$ 
varies as time $t$ evolves, as $\xi (t)$,
going from $\textsub\xi{min}$ (the initial position) to $\textsub\xi{max}$ (the destination site).
The electronic part is calculated somewhat similarly to Ref.~\cite{alkauskas2014first}, but adapted to cope with relatively large ionic movement in the migration process, as detailed below.

\subsubsection{Electronic Hamiltonian and wavefunctions}

We start from the Hamiltonian which depends on $\xi$, given by 
\[\underline{\textsub H e(\xi)}=\textsub V{II}(\xi)+ \underline {\textsub Te}+  \underline{\textsub V{Ie}(\xi)}+ \underline{\textsub V{ee}}.\]
Here, $T$ and $V$ denote kinetic and 
interaction energy operators, and ``I'' and ``e'' denote ions and electrons, respectively.
We use the underline to indicate 
many-electronic operators or wavefunctions to distinguish them from single-electronic ones. 
At any fixed ionic position, $\underline{\textsub H e(\xi)}$ is solved based on DFT, 
yielding the total ground state energy $\textsub Eg(\xi)$ as well as single-electron Kohn--Sham (KS) orbitals $\chi_i(\xi)$ with energies $\varepsilon_i(\xi)$. 
In the spirit of the KS scheme in DFT, the many-electron wavefunction is expressed by a Slater determinant consisting of KS orbitals
as $\underline{|\chi_{\{f\}}(\xi)\rangle}\equiv\mathcal A \underline{\left|\prod_i^{\{f\}}\chi_i(\xi)\right\rangle}$, where $\{f\}$ and $\mathcal A$ 
represent an electron filling of KS orbitals and anti-symetrization operator, respectively. 
Neglecting orbital relaxation when an electron is promoted to higher unoccupied KS orbitals, 
the electronic Hamiltonian is then written as
\begin{equation}\label{eq:he}\underline{\textsub H e(\xi)}\approx\sum_{\{f\}} E_{\{f\}}(\xi) \underline{|\chi_{\{f\}}(\xi)\rangle\langle\chi_{\{f\}}(\xi)|}.\end{equation}
Here, relying on Janak's theorem \cite{janak1978proof}, 
the energy is given as $E_{\{f\}}(\xi)=\textsub Eg(\xi)+\left(\sum_i^{\{f\}}\varepsilon_i(\xi)\right)-\left(\sum_i^{\{g\}}\varepsilon_i(\xi)\right)$, where $\{g\}$ denotes the orbital filling of the ground state.
On the other hand, 
since KS orbitals are eigen states of the single-electron KS Hamiltonian $\textsub H{KS}(\xi)$, we find 
\begin{equation}\label{eq:hks}\textsub H{KS}(\xi)=\sum_i \varepsilon_i(\xi) |\chi_i(\xi)\rangle\langle\chi_i(\xi)|.\end{equation}

Since all the quantities in \EQs{eq:he} and (\ref{eq:hks}) are available from the DFT program, $\underline{\textsub H e(\xi)}$ and $\textsub H{KS}(\xi)$ can then be constructed.
In the actual calculation, the PAW formalism is used. Consequently, we deal with the pseudo-wavefunctions but the operators are replaced by the PAW version as well, recovering real quantities.
The KS orbitals and energies at the $\Gamma$ point are used here to restore the correct band gap [though $\Gamma$ point is not used in the self-consistence field (SCF) calculation].

The carrier recombination process is governed by the highest occupied and the lowest unoccupied electron states since the orbital energies of those two states become close to each other during the Mg migration, thus causing the electron capture. The two orbitals in the present case are the defect and the CB KS orbitals. 
Other lower orbitals [the core and valence band (VB) states] are unlikely to participate in the electron-capture process during the migration, implying that adiabatic approximation can be applied to those 
states and electrons stay on the $\xi$-dependent KS orbitals $\chi(\xi)$ during the migration. For the defect and CB states, 
however,
adiabatic approximation no longer works to describe the carrier recombination process. 
We thus expand those states by using particular KS orbitals
at a fixed $\xi=\textsub\xi{max}$ [i.e., the final position], as in the static approach used in calculating capture coefficients for deep defects \cite{goguenheim1990theoretical,gutsche1982non,alkauskas2014first}. We have found that just including two orbitals, 
the CBM [$\textsub \chi c(\textsub\xi{max})$] and the defect [$\textsub \chi d(\textsub\xi{max})$], is sufficient in the present calculations. 
Noting that when $\xi\ne \textsub\xi{max}$, $\langle\chi_i(\textsub\xi{max})|\chi_v(\xi)\rangle\ne 0$ between $i\in \{\text c,\text d\}$ and any core or VB state $v$, 
in order to 
maximize the degrees of freedom of those basis orbitals, 
it is better to project out core and VB states from $\chi_i(\textsub\xi{max})$ as 
$\{|\chi_i(\textsub\xi{max})\rangle\}\mapsto\{|\psi_i(\xi)\rangle\}=\mathcal O\{\mathcal P_\xi|\chi_i(\textsub\xi{max})\rangle\}$, with $\mathcal O$ being the orthogonalization operator, and $\mathcal P_\xi=1-\sum_v^{\text{core \& VB}}|\chi_v(\xi)\rangle\langle\chi_v(\xi)|=\sum_v^{\text{defect \& CB}}|\chi_v(\xi)\rangle\langle\chi_v(\xi)|$. 
Hence, at any $\xi$, the wavefunction of the electron on the CB or the defect state is expanded using $\psi_i(\xi)$ ($i\in \{\text c,\text d\}$) 
and the total many-electron wavefunction is expressed by a 
linear combination of  
$\underline{|\tilde \psi_i(\xi)\rangle}\equiv\mathcal A \underline{\left|\left(\prod_v^{\text{core \& VB}}\chi_v(\xi)\right)\psi_i(\xi) \right\rangle}$.

For the matrix elements, we obtain
\begin{equation}\label{eq:mat}
\begin{split}
    \underline{\textsub He^{ij}(\xi)}&{}\equiv\underline{\langle\tilde \psi_i(\xi)| \textsub He(\xi)|\tilde \psi_j(\xi)\rangle}\\
    &{}=C(\xi)\delta_{ij}+\langle\psi_i(\xi)|\textsub H{KS}(\xi)|\psi_j(\xi)\rangle\\
    &{}=C(\xi)\delta_{ij}+\textsub H{KS}^{ij}(\xi),
\end{split}
\end{equation}
where $C(\xi)=\textsub Eg(\xi)-\textsub EV(\xi)$ with $\textsub EV$ denoting the energy of VBM 
being a function of $\xi$ but independent of $i$ and $j$ (also does not enter the final result).
It is found that only single-electronic quantities finally appear in many-electronic matrix elements.

\subsubsection{Electron capture probability}

As stated above, the electron capture mainly happens at the crossing point of the energy surfaces of two charge states.
The capture probability at the crossing can be obtained similarly to the Landau--Zener model \cite{landau1932theorie,zener1932non,henry1977nonradiative}.
Specifically, we consider the time-dependence of the electronic states.
The migration coordinate $\xi$ is now a function of time $t$, starting from $\xi=\textsub\xi{min}$ at $t=0$ and ending with $\xi=\textsub \xi{max}$ at a particular time $t=\tau$.
Let the total electronic wavefunction be
\[\begin{split}
    \underline{|\Psi(t)\rangle}&{}=\textsub \alpha c(t)\exp\left(\frac {-\ii}\hbar \int_0^t \underline{\textsub He^\text{cc}(\xi(t'))} \,\dd t'\right)\underline{|\textsub{\tilde \psi} c(\xi(t))\rangle}\\
    &{}+\textsub \alpha d(t)\exp\left(\frac {-\ii}\hbar \int_0^t \underline{\textsub He^\text{dd}(\xi(t'))} \,\dd t'\right)\underline{|\textsub{\tilde \psi} d(\xi(t))\rangle}.
\end{split}
\]
At $t=0$, the extra electron occupies CB and thus $\textsub \alpha c(0)=1$ and $\textsub \alpha d(0)=0$.
The Schr\"odinger equation $\ii\hbar \frac{\partial \underline {\Psi(t)}}{\partial t}=\underline{\textsub He(\xi(t))\Psi(t)}$ leads to
\begin{equation}\label{eq:schro}
\begin{split}
    \ii\hbar \frac{\dd\textsub \alpha d(t)}{\dd t}&{}=\textsub H{KS}^{\text{cd}}(\xi(t)) \exp\left(\frac {-\ii}\hbar \int_0^t \bigl(\textsub H{KS}^{\text{cc}}(\xi(t'))\right.\\
    &\left.\vphantom{\int_0^t}{}-\textsub H{KS}^{\text{dd}}(\xi(t'))\bigr) \,\dd t'\right) \textsub \alpha c(t).
\end{split}
\end{equation}
Here, we have used \EQ{eq:mat} and omitted terms with $\underline{\langle\textsub{\tilde \psi}c(\xi(t))|}\frac{\partial}{\partial t}\underline{|\textsub{\tilde \psi}d(\xi(t))\rangle}$ which are found to be small from our numerical calculations. 
The approximation to the lowest order of $\textsub H{KS}^{\text{cd}}$ is obtained by keeping $\textsub \alpha c(t)=1$ on the right-hand side of \EQ{eq:schro}, with the result
\begin{equation}\label{eq:intergral}
    \begin{split}
        \textsub \alpha d(t)&{}=\frac{-\ii}{\hbar}\int_{\textsub\xi{min}}^{\xi(t)}\textsub H{KS}^{\text{cd}}(\xi) \exp\left(\frac {-\ii}\hbar \int_{\textsub\xi{min}}^\xi \bigl(\textsub H{KS}^{\text{cc}}(\xi')\right.\\
        &\left.\vphantom{\int_0^\xi}{}-\textsub H{KS}^{\text{dd}}(\xi')\bigr) \frac{\dd \xi'}{\dot\xi'}\right)\frac{\dd \xi}{\dot\xi},
        \end{split}
\end{equation}
where $\dot\xi'=\left.\frac{\dd \xi(t)}{\dd t}\right|_{\xi=\xi'}$.
The exponential term in \EQ{eq:intergral} is highly oscillatory except near the crossing point of $\textsub H{KS}^{\text{cc}}(\xi)$ and $\textsub H{KS}^{\text{dd}}(\xi)$. Thus, main contribution in outer integral only comes from the region close to the crossing point, with $|\textsub \alpha d(t)|$ very small before the crossing and approximately constant after the crossing. Using stationary-phase approximation \cite{courant2008methods}, we can obtain the final capture probability 
\begin{equation}\label{eq:catP}p=|\textsub \alpha d(\tau)|^2\approx\frac{2\pi |\textsub H{KS}^{\text{cd}}(\xi_0)|^2}{\hbar  \left|\left.\dot\xi\frac\partial{\partial \xi}(\textsub H{KS}^{\text{cc}}-\textsub H{KS}^{\text{dd}})\right|_{\xi=\xi_0}\right|},
\end{equation}
where $\xi_0$ is the crossing point.

When the electron density is $n$, 
under the non-degenerate condition, the overall capture probability $P$ is then approximately 
\begin{equation}\label{eq:catPP}P= 1-(1-p)^{n\Omega}\approx 1-\ee^{-np\Omega},\end{equation}
with $\Omega$ the total volume and $p$ given by \EQ{eq:catP}
(the rightmost expression is valid when $p$ is small).

We note that when neglecting the approximation when deriving \EQ{eq:he}, the quantity $\textsub H{KS}^{\text{dd}}(\xi)-\textsub H{KS}^{\text{cc}}(\xi)$ should be equal to the total energy difference between two charge states.
Thus, the comparison between these two quantities can serve as a validation of the calculation. 

\subsection{Extrapolation of cell size}

In the actual calculation, the matrix element $\textsub H{KS}^{ij}$ is evaluated using finite-size supercells ($4 \times 4 \times 3$ in the present case), evaluated at charge state either before or after the electron capture (we use $1+$ in section \ref{sec:rr}). 
However, the finite-size supercell fails to describe the real CB state perturbed by the Mg interstitial
in an otherwise perfect crystal (aka. the scattering state) $\textsub \psi c$. 
Thus, the result must be extrapolated to supercell volume $\Omega\to\infty$.
In this limit, $\textsub H{KS}^{\text{cd}}(\xi_0)\to 0$ but $\Omega|\textsub H{KS}^{\text{cd}}(\xi_0)|^2$ approaches to a finite value.
Ref.~\cite{alkauskas2014first} has pointed out that this value can be evaluated using a simple model.
Let \[f\equiv\frac{\lim_{\Omega\to\infty}\Omega|\textsub H{KS}^{\text{cd}}(\Omega)|^2}{\textsub \Omega{cell}|\textsub H{KS}^{\text{cd}}(\text{cell})|^2}.\]
Here, $\Omega$ and $\textsub H{KS}^{\text{cd}}(\Omega)$ denote the volume of the supercell and the computed matrix element, respectively (note that the $\xi$ dependence has been omitted for the simplicity of the expression),
$\textsub H{KS}^{\text{cd}}(\text{cell})$ is the value calculated directly by DFT in the finite-size supercell with its volume $\textsub \Omega{cell}$. 
Because $\textsub H{KS}^{\text{cd}}$ is the matrix element between the well localized and almost $\Omega$-independent defect wavefunction $\textsub \psi d$ and extended and $\Omega$-dependent CB wavefunction $\textsub \psi c$,
we have $|\textsub H{KS}^{\text{cd}}(\Omega)|^2 \propto |\textsub \psi c^{(\Omega)} (\bm 0)|^2$, with $\textsub \psi c^{(\Omega)}$ denoting the CB wavefunction in cell $\Omega$ and assuming that the defect is located at the origin $\bm 0$. Hence,  
\begin{equation}\label{eq:fpsi}
    f=\frac{\lim_{\Omega\to\infty}\Omega|\textsub \psi c^{(\Omega)} (\bm 0)|^2}{\textsub \Omega{cell}|\textsub \psi c^{(\text{cell})} (\bm 0)|^2}.
\end{equation}
Both $\textsub \psi c^{(\Omega)}$ and $\textsub \psi c^{(\text{cell})}$ are then computed by effective mass approximation with $m^*=0.2\textsub me$ \cite{persson2001effective} and a model defect potential \cite{alkauskas2014first} 
\[u(r)=\frac{-Ze^2}{4\pi\varepsilon_0\varepsilon r}\erf(r/r_0)\exp(-r/\lambda),\]
where $Z$ is the defect charge used to calculate $\textsub H{KS}$ (1 in the current case), $\varepsilon=8.9$ is the dielectric constant of GaN, $r_0=\SI 2\angstrom$ is typical size of the defect, and $\lambda=\SI{60}{\angstrom}$ is the estimated Debye--H\"uckel screening length (has negligible effect on the final results).

When $\Omega\to\infty$, $\textsub \psi c^{(\Omega)}$ becomes the scattering state of potential $u(r)$.
We set incident energy relative to the CBM being equal to the average thermal kinetic energy ($\frac{\hbar^2 k^2}{2m^*}=\frac32 \textsub kBT$), corresponding to a thermal average of non-degenerate electrons on CB.
Specifically, we solve
\[\left[\frac{-\hbar^2\nabla^2}{2m^*}+u(r)\right]\textsub \psi c^{(\Omega)}(\bm r)=\frac{\hbar^2 k^2}{2m^*} \textsub \psi c^{(\Omega)}(\bm r)\]
using partial-wave analysis, with the boundary condition $\textsub \psi c^{(\Omega)}(\bm r)\sim \exp(\ii kz)/\sqrt{\Omega}$ for large $r$.
For $\textsub \psi c^{(\text{cell})}$, we put the potential $u(r)$ into the same periodic supercell as used in the DFT calculation, and solve the ground state wavefunction using planewave basis set.

For attractive defect potential, $f$ is usually larger than 1, meaning that \EQ{eq:catPP} underestimate the real capture probability. In the current case, we have found $f\approx 14$ at \SI{e3}{K}.

\section{Charge-state dependent migration of \uppercase{M}\lowercase{g} interstitial}

\begin{table}
    \caption{
Calculated migration barriers of the Mg interstitial with different XC functionals, GGA and HSE. Mg starts at as Mg$_{\rm O}$ with doubly positive charge state, then migrate to neighboring equivalent O site via the metastable interstitial complex \MgGaic{} with keeping its charge state (2+), with capturing a single electron (1+), or capturing 2 electrons (0). Calculations are performed with $2\times2\times2$ $k$ points and $\EF$ is taken to be CBM.}
    \label{tb:hp}
    \begin{tabular}{lll}
    \hline\hline
    Charge state  &  GGA (eV)  &  HSE (eV) \\ 
    \hline
    $2+$    & 2.25 & 2.23 \\
    \hline
    $1+$    & 1.66 & 1.65 \\
    \hline
    0 & 1.56 & 1.53 \\
    \hline
    \hline
    \end{tabular}
    \end{table}

\begin{figure}
    \includegraphics[scale=.63]{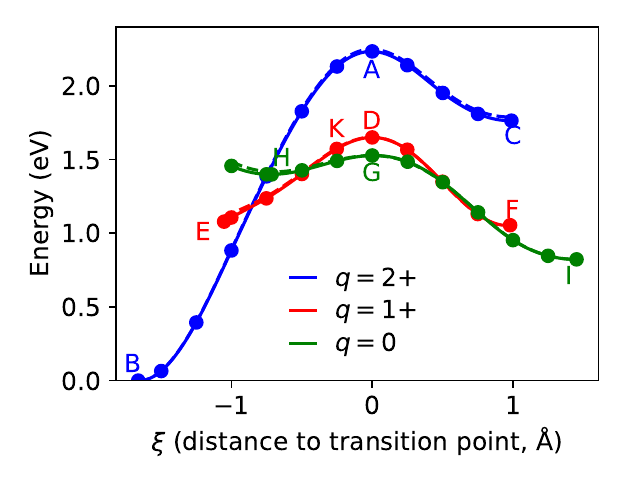}
    \caption{
The total energy profile along the migration pathway from the O site (left-end point B) to the midway geometry \MgGaic{} (right-end point C, F, or I) for each charge state.
The structures of points A$\sim$K are shown in \FIG{fg:path} (c). Note that the same $\xi$ of different chare states does not refer to the same atomic structure. The solid and dashed lines show the result calculated using $2\times2\times2$ and single $\bm k$ points, respectively. Red and green dashed lines are shifted 
downwards by \SI{0.05}{eV} to match the solid lines.
}\label{fg:qf}
\end{figure}

\begin{figure}
    \includegraphics[scale=.63]{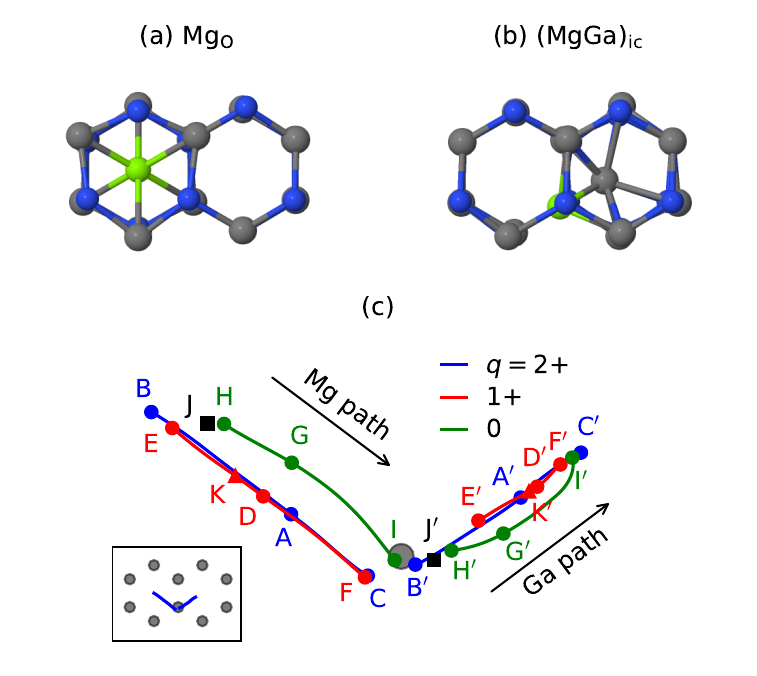}
    \caption{Schematic illustration of the atomic structure of Mg\ts O \textsf{(a)} and \MgGaic{} \textsf{(b)}, respectively, viewed along the $\langle 0001 \rangle$ axis. Blue, gray and green balls represent N, Ga and Mg, respectively.
    \textsf{(c)} The trajectory of the migrating Mg atom and the nearby Ga atom, projected on the $(0001)$ plane, in the Mg\ts O $\to$ \MgGaic{} migration shown in \FIG{fg:qf}. The energies of points A$\sim$K (for Mg, and the corresponding Ga positions are marked by A$'$$\sim$K$'$) are shown in \FIG{fg:qf}.
    The projected positions B$\to$A$\to$C for the $2+$ 
migration, E$\to$K$\to$D$\to$F for the $1+$ migration, and H$\to$G$\to$I for the $2+$ migration are 
depicted by blue, red and green dots, respectively, both for the Mg and Ga positions. 
The projected positions of the geometry J (see text) 
at which the total-energy between $1+$ and neutral crosses with minimum value are shown by black squares.  
The inset is a zoom-out view in which gray dots denote the lattice sites (at ideal positions), corresponding to the blue and gray balls shown in \FIG{fg:path} (a) and (b), and blue lines represent the migration trajectories of the Mg and Ga atoms.
}\label{fg:path}
\end{figure}

In our recent Letter \cite{zhao2024first}, we have explored stable and metastable structures of Mg interstitial atom in wurtzite GaN and determined its migration pathways and corresponding energy barriers with various charge states by GGA in DFT: In addition to doubly positive Mg at O site Mg$_{\text O}^{2+}$ [a part of the geometry shown in \FIG{fg:path} (a)] and at tetrahedral site Mg$_{\text T}^{2+}$ which were discussed earlier \cite{miceli2017migration}, we have found new metastable geometries, the complex of Mg and Ga interstitials \MgGaic{} [\FIG{fg:path} (b)] and a pair of a substitutional Mg and a Ga interstitial Mg$_{\text{Ga}}\text{Ga}_{\text i}$. More interestingly, those metastable geometries are found to capture one or two electrons depending on the $\EF$ position and become $1+$ or neutral charge states, respectively, opening a possibility of recombination-enhanced/retarded migration of the Mg interstitial. We have indeed found that the most stable Mg$_{\text O}^{2+}$ migrates to the neighboring O site via those metastable geometries with $1+$ or neutral charge states with the reduced migration barriers compared with the simple migration between the two equivalent O sites with $2+$ charge state. It is found that the amount of the reduction in the barrier is the largest in case of the migration via \MgGaic. The total energy profile along such migration pathway from the O site via \MgGaic{} (to another O site) obtained in the present HSE calculations is shown in \FIG{fg:qf} and the resultant migration barriers with ($1+$ and 0) and without ($2+$) electron capture are tabulated in \TBL{tb:hp}. The corresponding result with GGA calculations were shown in Figs.~4(a) and (b) in Ref.~\cite{zhao2024first}.

In obtaining the total-energy profiles as in \FIG{fg:qf}, 
we have first determined the transition states (the saddle point on the migration pathway), as marked by A, D and G points, for charge states $2+$, $1+$ and 0, respectively, 
by the CI-NEB method.
The formation energies 
of those transition-state geometries
relative to Mg\ts O$^{2+}$, i.e., the migration energy barrier, is found to be 2.23, 1.65 and \SI{1.53}{eV}, respectively, 
when $\EF$ is located at CBM.
Using GGA for structural optimization, similar values can be obtained 
as in \TBL{tb:hp}. Hence, we reasonably argue that the GGA and the HSE approximations provide essentially identical results on the migration barriers. It is still of note that the GGA 
values here are slightly higher than the values reported in our previous Letter. We have found that this is mainly the result of different lattice constants used in the calculation (we previously used the lattice constant optimized by GGA).

The other points in \FIG{fg:qf} are obtained by descending from the saddle points.
From the transition states to the \MgGaic{} geometry, the destination point \MgGaic{} marked by C, F and I in \FIG{fg:qf} are reached for all charge states.
However, going along the opposite direction toward Mg\ts O,
only $2+$ charge state can reach the initial B point (since Mg\ts O only has $2+$ charge state). 
For charge state $1+$, the pathway continues to E point and then disappears since the electron trapped by the defect is released to the conduction band. 
For neutral charge state, we instead arrive at another local minimum marked by H in \FIG{fg:qf} that is stable only at neutral charge state.
The energy will increase when going further away from the point H (the left end point of green line in \FIG{fg:qf}) before the electron is finally released into the CB.

It is noteworthy to recall that the migration pathways are along valleys and with saddle points in the total-energy landscape on the multi-dimensional space of atomic coordinates. The pathways are intrinsically distinct for different charge states. The total-energy profiles in \FIG{fg:qf} are just projections of the pathways on single migration coordinates $\xi$ obtained by the NEB calculations, which are not identical among different charge states. This situation is well demonstrated in \FIG{fg:path} (c). The migration pathway from Mg\ts O to \MgGaic, for which the energy profile is shown in  \FIG{fg:qf}, of course involves a complex atomic-structure variation, but in a simple-minded picture the Mg atom at the O site pushes the nearby Ga atom at the lattice site to the interstitial site and forms \MgGaic{} complex. \FIG{fg:path} (c) shows the position of the Mg atom and the nearby Ga atom on a particular $(0001)$ plane. In the migration for the $2+$ and $1+$ species, the trajectory of Mg and the Ga are close, suggesting that the electron capture takes place near the crosspoint of the total-energy profiles for the $2+$ and $1+$ in \FIG{fg:qf}, without additional total-energy barrier. However, the migration pathway for the neutral charge state is substantially different from those for the $2+$ and $1+$ charge states, as depicted by the green lines in \FIG{fg:path} (c) and the appearance of a local minimum on the energy surface as discussed above, indicating a necessity of further geometry exploration.

Because the second electron capture occurs when the formation energies of $1+$ and neutral states cross,
the minimum energy among all such crossing points corresponds to the energy barrier of electron capture.
To find its value, we have performed a structural optimization with the constraint that the formation energies between $q=1+$ and 0 are equal
(with $\EF$ being to be CBM), and then found the J point as shown in \FIG{fg:path} (c), with an energy of \SI{1.55}{eV} relative to Mg\ts O$^{2+}$, higher than the migration barrier of neutral charge state (\SI{1.53}{eV} represented by the saddle point G in \FIG{fg:qf}).
This indicates that starting as Mg\ts O$^{2+}$, a minimum of \SI{1.55}{eV} is required to move to J (with the first electron capture on the way), where the second electron is captured, and then continue to migrate at neutral charge state (actually without additional energy barriers more than this \SI{1.55}{eV}).
However, it should be noted that this value is lower than the migration barrier of $q=1+$, therefore, the second capture is still beneficial.

\section{Recombination rates}\label{sec:rr}

\begin{figure}
    \includegraphics[scale=.63]{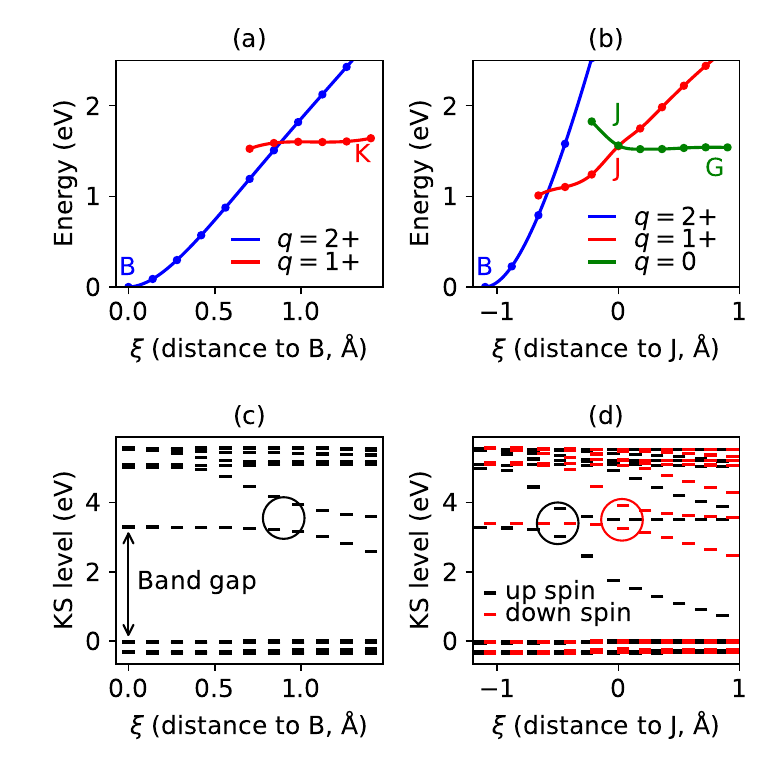}\\
    \caption{%
        \textsf{(a)} The total energies of charge state $2+$ and $1+$ along the BK path. Unlike \FIG{fg:qf}, the same $\xi$ here does correspond to the same atomic structure. 
        \textsf{(b)} The same as (a), but for BJG path.
        \textsf{(c)} The KS levels calculated at $q=1+$ along the BK path that are used to evaluate the recombination rate. 
        Only levels for up spin are shown since the down-spin electrons do not participate the capture.
        Electrons are filled up to the lowest level above the band gap. The KS levels exhibit an avoided crossing when total energies
        crosses, as marked by the circle. 
        \textsf{(d)} The same as (b), but for BJG path. Here, KS levels calculated at $q=1+$ are shown for both up-spin and down-spin electrons, which participate the first and second captures, respectively.
}\label{fg:qv}
\end{figure}

\begin{figure}
    \includegraphics[scale=.63]{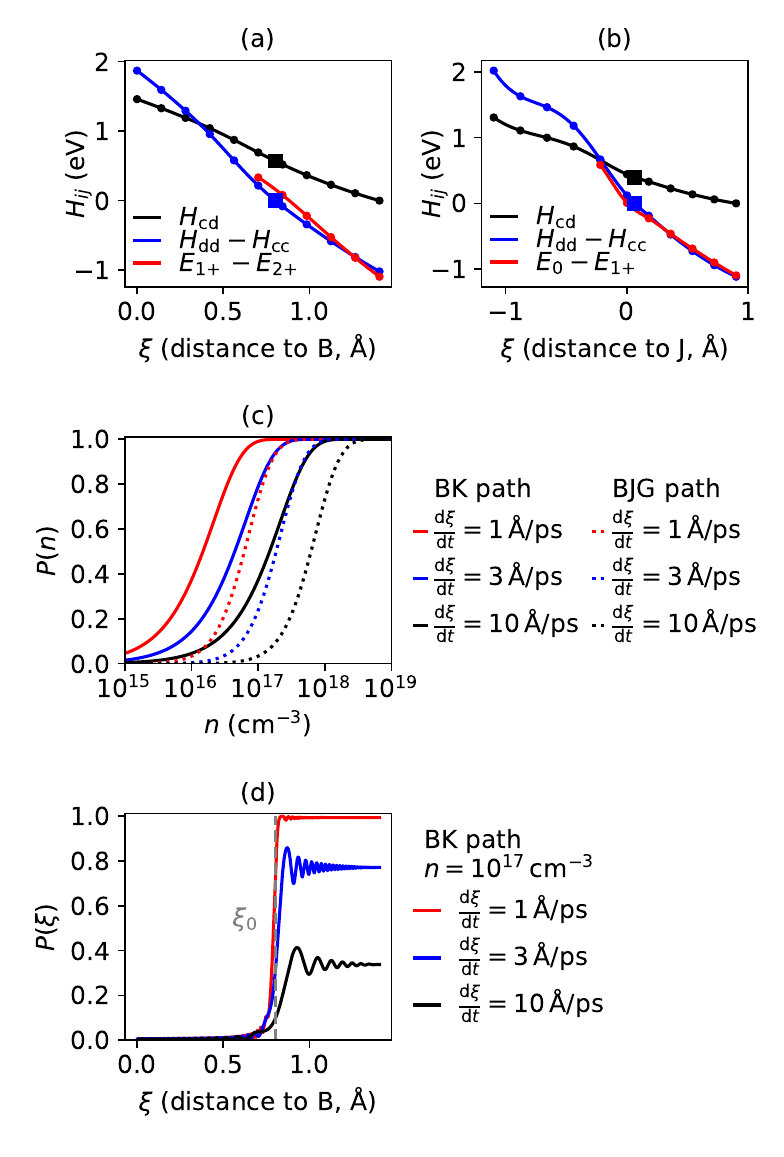}\\
    \caption{\textsf{(a)} and \textsf{(b)} The matrix elements for the electron capture on BK and BJG pathways (the second capture for the latter), respectively. 
    Blue and red lines correspond to 
the total-energy difference between the two distinct charge states (see text) with 
its zero-value indicating the cross point between the two charge states (blue square). 
The matrix element $\textsub H{cd}$ for the electron capture at the cross point is 
emphasized by the black square.
    \textsf{(c)} The overall capture probability as a function of electron density $n$ and migration speed $\dd\xi/\dd t$. 
    \textsf{(d)} The capture probability at each $\xi$ along the BK pathway. The vertical dashed line depicts the position of $\xi_0$.
    }\label{fg:cap}
\end{figure}

Having found out the energies required for electron capture, we now proceed to calculate its probability during the migration. In section \ref{sec:cap}, we have presented the method for such calculation. The key quantities are
$\textsub H{KS}^{\text{dd}}-\textsub H{KS}^{\text{cc}}$ which 
corresponds to the total energy difference between the two charge states, and $\textsub H{KS}^{\text{cd}}$ which is the coupling strength between CB and defect state. Here we discuss the recombination rates in the two migration pathways
summarized in the last section.
In the first one, neutral charge state is not considered. The Mg starts from Mg\ts O$^{2+}$ (B point), captures one electron, passes D and reaches \MgGaic $^{1+}$ (F point).
We consider the electron capture on the straight path connecting B and K, a point on $q=1+$ pathway slightly prior to D.
This point is chosen because we have found that connecting BD instead will raise the migration barrier.
From K, the migration can proceed with the fixed charge state of $1+$, with the energy barrier being \SI{1.65}{eV} at D point as discussed above.
For the second case, double electron captures on the path B$\to$J$\to$G connected by two straight lines are considered.
Passing through the J point requires \SI{1.55}{eV}, which is also sufficient to then reach \MgGaic $^0$ (I point) via G.

\FIG{fg:qv} (a) and (c) show the total energies calculated with $q={2+}/{1+}$ on the BK path and KS levels calculated at $q=1+$ (the electronic Hamiltonian is subsequently calculated at this charge state) \footnote{Strictly speaking, near the B point, Mg is at charge state $2+$ with the extra electron occupying CB. While such state is unphysical in super-cell calculation, the recombination probability is dominated by the matrix element at the crossing point, which can be accurately evaluated as stated in subsection \ref{sec:dft}.}, respectively.
The migration starts with $q=2+$ at $\xi=0$ (B point).
Near $\xi\approx\SI{0.8}{\angstrom}$, the crossing of the total energies and the KS levels is encountered (the KS levels actually only exhibit an avoided crossing due to coupling between CB and defect states), enabling the electron capture.
After the capture, the energy profile of $q=2+$ [blue line in \FIG{fg:qv} (a)] is transferred to that of $1+$ (red line), and the migration continues with reduced energy barrier. 
The matrix elements for calculating the capture probability at the crossing is shown in \FIG{fg:cap} (a).
We first observe the matrix element $\textsub H{KS}^{\text{dd}}-\textsub H{KS}^{\text{cc}}$ (blue line) well agree with the difference in total energies (red line), showing the validity of the calculation method.
Reading off the quantities at the crossing point $\xi_0$ [when $\textsub H{KS}^{\text{cc}}-\textsub H{KS}^{\text{dd}}=0$ as indicated by the blue square in in \FIG{fg:cap} (a)], i.e.,
$\textsub H{KS}^{\text{cd}}(\xi_0)$ (black square) and $\left.\frac\partial{\partial \xi}(\textsub H{KS}^{\text{cc}}-\textsub H{KS}^{\text{dd}})\right|_{\xi=\xi_0}$ (the slope of blue curve at the blue square), we obtain the capture probability during the whole migration using \EQs{eq:catP} and (\ref{eq:catPP}) at different migration speed $\dd\xi/\dd t$ and electron densities. The results are shown by solid lines in \FIG{fg:cap} (c).
Considering that the migration typically proceeds within a time scale of \si{ps} order and a distance of \si{\angstrom} order, we conclude that high capture probability can be achieved for electron density $n>\SI{e17}{cm^{-3}}$, 
and the recombination-enhanced migration is expected to be significant under this condition.
Additionally, the capture probability at each $\xi$ along the migration pathway [\EQ{eq:catP} only corresponds the final capture probability]
can be computed by numerically integrating \EQ{eq:intergral}. The results are shown in the results given in \FIG{fg:cap} (d).
It is evident that the capture probability behaves like a step function at the crossing point $\xi_0$, which confirms our physical pictures in subsection \ref{sec:cap}: The capture probability is nearly constant before and after the crossing where the electronic wavefunctions nearly adiabatically follows the motion of ions; however, at the crossing point, the electron capture occurs quickly and the probability reaches a high value in just a short interval of $\xi$.

On the second pathway BJG, as shown in \FIG{fg:qv} (b) and (d), firstly an up-spin electron is captured near the crossing at $\xi=\SI{-0.5}{\angstrom}$.
After the capture, the migration proceeds with $q=1+$, and then the second down-spin electron is captured near the crossing at J point, with charge state becoming neutral.
The whole probability of capturing two electrons is the product of the two individual captures, which can be calculated just the same way as the BK path.
We have found that the two individual captures also exhibit similar behavior as in the previous case [the matrix elements for the second capture are shown in \FIG{fg:cap} (b)], and the probability of double capture is depicted by the dotted lines in \FIG{fg:cap} (c), suggesting that under a higher electron density $10^{17}\sim\SI{e18}{cm^{-3}}$, this double capture process also becomes feasible.

To complete the migration step, the Mg then migrates from \MgGaic{} to the next O site, with one or two electrons released on the pathway. 
The calculation of the emission probability is very similar to the capture probability,
except that, unlike the capture process which requires occupied CB states to provide electrons, for emission process, the electron can be ejected to any state on CB.
Thus, in \EQ{eq:catPP}, the electron density $n$ is to be replaced by the effective density of states of CB, $\textsub NC$.
Since generally $\textsub NC>n$, the probability of electron emission is expected to be high and not a bottleneck for recombination-enhanced migration.

\section{Conclusion}

We have determined stable structures and migration pathways with accurate HSE 
approximation to the exchange-correlation energy, and also computed 
recombination rates using the obtained energy spectrum and wavefunctions.
We have found that the migration between O sites via \MgGaic{}
shows 
the lowest migration energy in which one or two electrons are captured 
during the migration, the most stable charge state of $2+$ changes to $1+$ or 
neutral, and by this recombination of carriers the migration barrier is significantly 
reduced.  When $\EF$ is high in the gap, starting from Mg\ts O$^{2+}$, Mg captures an electron becoming the $1+$ charge 
state and overcomes the barrier of 1.65 eV, much reduced from 2.23 eV in case of 
the migration with the $2+$ charges state kept. Moreover, further electron capture 
is realized accompanied by substantial structural relaxation, thus Mg becoming 
neutral. Detailed HSE calculations for this second capture show that the migration 
barrier is 1.55 eV, thus clarifying the important role of the carrier recombination 
in Mg migration in GaN. These theoretical findings are corroborated by the present 
quantitative calculations of recombination rates based on our DFT-obtained Hamiltonian matrix elements and Landau--Zener model, 
which reveals that the time-scale of the recombination is in or under 
the timescale of the migration with typical electron density. This clearly indicates that the carrier recombination during migration is highly probable.  

\begin{acknowledgments}
This work is supported by the MEXT programs ``Creation of innovative core technology for power electronics'' (Grant Number JPJ009777),
as well as Grant-in-Aid (Kakenhi) for Japan Society for the Promotion of Science (JSPS) Fellows No. 22KJ0579. 
The computation in this work has been done using the facilities of the Supercomputer Center, the Institute for Solid State Physics, The University of Tokyo.
\end{acknowledgments}

\bibliography{rc.bib}

\end{document}